# Nonlinearity enabled higher-dimensional exceptional topology

Kai Bai,[1] Meng Xiao,[1, 2*]

[1]Key Laboratory of Artificial Micro- and Nano-structures of Ministry of Education and School of Physics and Technology, Wuhan University, Wuhan 430072, China

[2]Wuhan Institute of Quantum Technology, Wuhan 430206, China

Corresponding Email [*]: phmxiao@whu.edu.cn

**Abstract:**

The role of nonlinearity on topology has been investigated extensively in Hermitian systems, while nonlinearity has only been used as a tuning knob in a PT symmetric non-Hermitian system. Here, in our work, we show that nonlinearity plays a crucial role in forming topological singularities of non-Hermitian systems. We provide a simple and intuitive example by demonstrating with both theory and circuit experiments an exceptional nexus (EX), a higher-order exceptional point with a hybrid topological invariant (HTI), within only two coupled resonators with the aid of nonlinear gain. Phase rigidities are constructed to confirm the HTI in our nonlinear system, and the anisotropic critical behavior of the eigenspectra is verified with experiments. Our findings lead to advances in the fundamental understanding of the peculiar topology of nonlinear non-Hermitian systems, possibly opening new avenues for applications.

**Introduction**

The physical dimension of a system is of vital importance in topological physics, and a higher dimension enables diverse and complex topology. The interplay of the dimension and topology of a non-Hermitian system is even more fascinating, and even a one-dimensional single band can exhibit nontrivial topology as the complex plane is introduced [1]. The recent advances in non-Hermitian physics are underlain by the "exceptional" constituents [2–15]. There are exceptional points (EPs) [16,17], rings [18,19], surfaces [20], bulk exceptional arcs (EAs) [21], and exceptional nexuses (EXs) [22] as a manifestation of different physical dimensions. Higher physical dimensions enable the forming of higher-order EPs which then lead to enhancement of light-matter interactions[23,24], sensing [25], and mechanical damping and the spring stiffness in optomechanics [26]. Of all the higher-order EPs, EXs stand out, as they not only possess a hybrid topological invariant (HTI) [22] which exhibits distinct winding numbers along different cyclic paths, but also are joints of cusp singularities of multiple EAs. EXs were originally believed to be realized only within three-dimensional (3D) systems with three independent knobs. Here, we provide a new approach to realize higher dimensional EPs with the aid of nonlinearity and demonstrate that a 3D EX can be achieved even within 2 cavities.

The integration of nonlinearity with topology in Hermitian systems brings fruitful novel phenomena [27–30], while research on the exceptional singularities of nonlinear non-Hermitian topological physics is rare [31-35]. In stark contrast, nonlinearity is natural in non-Hermitian systems especially when gain is introduced. From the energy consideration, one needs to consider a more realistic nonlinear saturable gain which then leads to beneficial applications such as wireless energy transfer [36] and optical bistability [37]. To date, the vast majority of works on non-Hermitian systems have been restricted to the linear region. Our work provides a simple and intuitive example and shows how the nonlinear saturable gain can introduce a hidden dimension that further lead to an EX. Different from previous works [27–37], such physics discussed here with an additional dimension cannot be captured within the linear region or its extension with weak nonlinearity. Our work thus points to the possibility of exploring higher-dimensional ($\geq 3$) physics in a lower-dimensional ($= 2$) system with the aid of nonlinearity. Moreover, the exquisite dimension correlation in our system offers us unique advantages in utilizing the exceptional features of non-Hermitian systems. A higher-order EP such as an EX can greatly enhance the sensitivity of the system [25,38,39]. However, to reach the subtle EX conditions, time-consuming tedious parameter tuning in a hyperdimensional space must be performed. Here, this seeming incompatibility can be naturally reconciliated: the EX condition can be reached by changing only very limited parameters (even fewer than those needed to form an order-2 EP). Meanwhile, the EXs herein are more robust against noise than their counterparts in linear systems thanks to the feedback mechanism of nonlinear saturable gain.

**General theoretical analysis**

Our system consists of two coupled resonators with nonlinear gain (red) and loss (blue) as sketched in the upper panel of Fig. 1(a). The harmonic mode with frequency $\omega$ satisfies the self-consistent nonlinear equation [36,37]

$$\begin{pmatrix} \omega_A + ig(|\psi_A|) & \kappa \\ \kappa & \omega_B - il \end{pmatrix} \begin{pmatrix} \psi_A \\ \psi_B \end{pmatrix} = \omega \begin{pmatrix} \psi_A \\ \psi_B \end{pmatrix}, \quad (1)$$

where $\omega_A$ ($\omega_B$) is the resonant frequency of the left (right) resonator, $\psi_{A,B}$ represents the corresponding field amplitude defined such that $|\psi_{A,B}|^2$ is the energy stored in each resonator, $\kappa$ denotes the strength of coupling, $l$ denotes the loss, and $g$ represents the gain, which is assumed to depend on $|\psi_A|$. The complex secular equation can be decomposed into real and imaginary parts as

$$(\omega - \omega_A)(\omega - \omega_B)^2 + (\omega - \omega_A)l^2 - \kappa^2(\omega - \omega_B) = 0, \quad (2)$$

$$g_s = \frac{(\omega_A - \omega)}{(\omega_B - \omega)} l. \quad (3)$$

Here noting that the magnitude of gain $g$ depends on the field amplitude, and thus Eq. (3) being satisfied at $g = g_s$ can be achieved by adjusting the wave function. Figure 1(b) shows the real $\omega$ solution of Eq. (2) versus loss $l$ and detuning $\Delta_\omega \equiv \omega_A - \omega_B$, where we set $\kappa = 1$ and $\omega_B = 0$. Here, for simplicity, all the parameters are normalized by $\kappa$ and become dimensionless. At zero detuning, there are three states when $l < \kappa$, as denoted by the bold (light) red and blue lines, and these three states coalesce into one at $l = \kappa$, as marked by the black star. Meanwhile, over a finite detuning range, these three states are preserved until two of them coalesce at the yellow lines. Note that here, the specific form of gain is irrelevant for obtaining Fig. 1(b) provided that Eq. (3) can be satisfied.

Previous studies show that the system Hamiltonian becomes defective at EPs whereat two or more eigenvalues and their corresponding eigenvectors coalesce into one [40]. The coalescence of states at the black star and yellow lines is reminiscent of the physics at the EPs and EAs in non-Hermitian systems. Looking into the details, Figs. 2(a)-(d) plot Re[$\omega$] for both the real (solid lines) and complex (dotted lines) solutions of Eq. (2) for different detuning and loss values. The solid lines in Fig. 2(a) for $l = 0.7$ show a typical bistable scenario. The red and blue solid lines correspond to stable and unstable states, respectively, based on computations of the Lyapunov exponents (See Supplementary Materials Sec. 1.). The red and blue solid lines coalesce at the two yellow points in Fig. 2(a) where only one state remains. We focus on the lower coalescence point ($\Delta_\omega = 0.22$) and plot Re[$\omega$] as a function of loss in Fig. 2(b), wherein similarly two states coalesce into one while the higher red line remains. Meanwhile, we find that both Re[$\omega$] and Im[$\omega$] approach the corresponding coalescence point as $\sqrt{\epsilon}$, with $\epsilon$ being the perturbation on either the detuning or loss (see Supplementary Materials Sec. 2), which thus indicates that the coalescence point is an order-2 EP (Two eigenstates coalesce into one). As a straightforward extension, the two yellow lines in Fig. 1(b) should be arcs of EPs, i.e., EAs. These two EAs approach each other with increasing $l$ and eventually merge at $l = 1$, where only one state remains at the coalescence point marked by the black star, as shown in Figs. 2(c) and (d).

The critical scaling behavior can be obtained by analyzing Eq. (2), the secular equation, which itself is a cubic function of $\omega$. This is reminiscent of the secular equation of systems possessing three resonators [22]. At $\Delta_\omega = 0$, Eq. (2) is simplified as

$$\omega = \pm\sqrt{1-l^2}, 0; \tag{4}$$

and at $l = 1$, Eq. (2) reduces to

$$\omega^3 - \omega^2 \Delta_\omega - \Delta_\omega = 0, \tag{5}$$

where compared to the third term, the second term can be ignored. Interestingly, Re[$\omega$] and Im[$\omega$] scale as $\sqrt[3]{\Delta_\omega}$ and $\sqrt{l-1}$, respectively, when approaching $\Delta_\omega = 0$ and $l = 1$, indicating an anisotropic order-3 EP [41], which was originally believed to be possible only in three- or even higher-dimensional systems. In other words, we have realized higher-dimensional ($\geq 3$) exceptional physics within two coupled resonators with the aid of nonlinearity. Note that, though Eqs. (1) and (2) had been discussed before in the circumstance of wireless energy transfer [36] and optical bistability [37], the parameter regime containing such an order-3 EP was not probed before.

To understand the underlying physics, we embed our two-resonance nonlinear system into a specific

$PT$ symmetric three-resonance linear Hamiltonian $H_{3d}$. This can be done as $H_{3d}$ shares the same manifold (eigenvalues and eigenvectors) as our two-resonator nonlinear system except for an auxiliary "neutral" site as sketched in the lower panel of Fig. 1(a). (The proof is provided in Supplementary Materials Sec. 3.). Such a map can be derived as follows. Let $\{\omega_i | (\psi_{i,A}, \psi_{i,B})^T\}$ with $i = 1, 2, 3$ and superscript T short for transpose represent the three eigenvalues and eigenvectors of Eq. (1). Except for the parameters exhibiting EPs, there are always three solutions since Eq. (2) is cubic. The eigenvectors can be easily determined to be not biorthogonal, and the space expanded by the eigenvectors is overcomplete. To construct the linear Hamiltonian $H_{3d}$ and restore the biorthogonality, we extend the eigenstates by including an auxiliary "neutral" site and the new vectors are $|\phi_i\rangle = (\alpha_i \psi_{i,A}, \alpha_i \psi_{i,B}, \psi_{i,N})^T$, with $\alpha_i \in \mathbb{C}$ and $\psi_{i,N}$ representing the wave amplitude inside the new site. The values of $\alpha_i$ and $\psi_{i,N}$ can be uniquely determined by assuming that $|\phi_i\rangle$ satisfies the biorthogonal and normalized conditions. $|\phi_i\rangle$ is the eigenvector of a three-resonance linear Hamiltonian $H_{3d} = \sum_{i=1}^{3} \omega_i |\phi_i\rangle\langle\phi_i|$, where $\langle\phi_i|$ is the left eigenvector of $|\phi_i\rangle$ following the biorthonormal condition. Such a Hamiltonian can be proven to be $PT$ symmetric, and the $PT$ operator is provided in Supplementary Materials Sec. 3. We emphasize that $H_{3d}$ defined above is unique for the corresponding nonlinear two-resonator system, and except for the EPs, such a construction works for any choice of loss and detuning.

With the map established, the topology of the two-resonance nonlinear system can be revealed with $H_{3d}$. Emerging through the cyclic evolutions of eigenstates, the topology of the non-Hermitian eigenspace is also tied to the splitting of eigenstates [42-44] in the vicinity of EPs. Therefore, it can be characterized using phase rigidity which is defined as $r_i = (\langle\phi_i|\phi_i\rangle)^{-1}$ [22,41,44,45]. Figures 2(e)-(h) show the phase rigidities along the loss and detuning direction, where we can see that the phase rigidities vanish at the EPs. A more detailed study shows that the critical exponents are 1/2 in Figs. 2(e), 2(f) and 2(h) and 1/3 in Fig. 2(g) (Supplementary Materials Sec. 3), and hence, the EPs along the EAs are all normal order-2 EPs except for the anisotropic order-3 EP at the cusp singularity. This order-3 EP exhibits different critical exponents of phase rigidity along the loss and detuning axes, i.e., exhibits an HTI, and it is thus an EX [22].

A higher-order EP such as an EX can magnificently enhance the system's sensitivity due to the splitting of eigenfrequencies at the EP [25,38,39]. Encircling around the EPs in the parameter space unavoidably introduces a non-adiabatic process for both the linear and the nonlinear systems [37], which may disturb the sensitivity enhancement. However, the most severe impediment to practical application is the tediously detailed tuning in hyperdimensional parameter space. Here, our system offers a way out of this dilemma using the EX point achieved with nonlinear saturable gain. In our system, only the states represented by the solid red lines in Figs. 2(a)-(d) or equivalently the red surface in Fig. 1(b) are dynamically stable, i.e., can be experimentally probed. Any other excitations will fall back into one of the stable states within a short period considering that the gain is automatically adjusted with the wave amplitude. In contrast, the exceptional features such as Re[$\omega$] scaling as $\sqrt[3]{\Delta_\omega}$ associated with the EX are still preserved for the stable states. Intriguingly, such an EX point is reached with even fewer constraints or tuning parameters than for an order-2 EP, wherein, in addition to two tuning parameters, one needs additional efforts to enforce *PT* symmetry.

**Experimental verification**

The above discussed physics is general, and here, we realize it with a circuit system, as shown in Fig. 3(a). The system consists of two LC resonators coupled by a capacitor $C_c$. The LC resonator on the right-hand side is lossy, while the resonator on the left-hand side exhibits saturable gain realized through an effective negative resistor (Supplementary Materials Sec. 4). Voltages $V_A$ and $V_B$ represent the wave amplitudes inside the left and right resonators, respectively. The Kirchhoff equations that describe the dynamics of such a circuit can be mapped into the coupled mode equation, i.e., Eq. (1) by assuming $C_c \ll C$. Under this mapping, the resonance frequency, coupling, loss and gain are given by $\omega_{A,B} = 1/\sqrt{L_{A,B}C}$ , $\kappa = \omega_B C_c/2C$ , $l = 1/2R_B C$ and $g[V_A] = R_D[V_A]/2\, R_g R_1 C$, respectively (Supplementary Materials Sec. 5). $g[V_A]$ decreases with an increasing $V_A$, as $R_D[V_A]$ is a monotonic decreasing function of $V_A$ over the range of interest (Supplementary Materials Sec. 4). Figure 3(b) shows the experimental setup, where the oscilloscope measures the resonance frequency and the amplitudes of $V_A$ and $V_B$, the DC power supplies power for the amplifier, and the waveform generator is used to excite the selected mode. We also add a homemade variable inductor for fine tuning of $L_A$ and variable resistors to control the loss and detuning. Details of the circuit elements on the PCB can be found in Supplementary Materials Sec. 6

Figures 3(c)-(f) show the measured resonance frequencies for various detuning and loss values for the corresponding cases studied in Figs. 2(a)-(d), respectively. The open circles and diamonds are measured resonance frequencies, and the red and blue solid lines represent the stable and unstable steady states of the corresponding Kirchhoff equations, respectively. The unstable steady states cannot be reached experimentally, and the two bistable states can be selectively excited by a "kicking" process with an external waveform generator (Supplementary Materials Sec. 7). Figures 3(g)-(j) show the critical behavior near the EAs and EX. The slopes are fit to 1/2 for points on EAs along both the detuning and loss axes, and the slopes for the EX are 1/3 along the detuning axes and 1/2 along the loss axes which verifies the eigenvalue anisotropy of the EX. Meanwhile, we also measured the ratio of the voltages on both resonators, which also agrees almost perfectly with the simulations (Supplementary Materials Sec. 8). In our presenting experimental setup, we cannot verify the HTI due to the experimentally inaccessible unstable states, however our work already demonstrates the sensitivity enhancement due to the EX within two coupled resonators. Sensitivity enhancement in the vicinity of EPs had also been discussed in terms of quantum metrology [46–51]. Here we focus on the classical region where the resources are unlimited and thermal noise is not an issue (especially for our circuit system). Although only the circuit system is observed, our result can be generalized to diverse classical systems, ranging from photonics [52], mechanics [53], and acoustics [54] to active matter [55]. And, especially in optics, with the rapid development of metastable lasers [56,57] and nonlinear resonators [58], the possibility is even more conceivable.

**Conclusions**

In summary, we show with theory and experiments that an EX with an HTI can be realized within a two-resonance non-Hermitian system by incorporating nonlinear saturable gain. The nonlinear gain introduces bistable steady states and another unstable state that effectively extends the dimension of the system and enables the investigation of higher-dimensional non-Hermitian topology. Our work shows that the unstable steady states, which were generally considered to be irrelevant to the dynamics, have definite contributions to the exceptional features and topology of the system. The possibility of exploring even more fascinating "exceptional" topology in simple nonlinear non-Hermitian systems is thus conceivable. Highlighting the fundamental understanding of nonlinear non-Hermitian systems, topology and dimension, our findings also constitute a major advance to realize a higher-order EP sensor in practice.

**Acknowledgments**

The authors thank Shu-Chen Zhang, Tian-Rui Liu, Liang Fang, Xin Lv, Jia-Zheng Li and Hao Wang for help. We also thank Dr. Chong Chen, Prof. Kun Ding and Prof. Shanhui Fan for helpful discussion during the revision of this manuscript. This work is supported by the National Natural Science Foundation of China (Grant Nos. 11904264) M. X. is also supported by the startup funding of Wuhan University.

**Author information**

Correspondence and requests for materials should be addressed to M.X. (phmxiao@whu.edu.cn);

# Figures

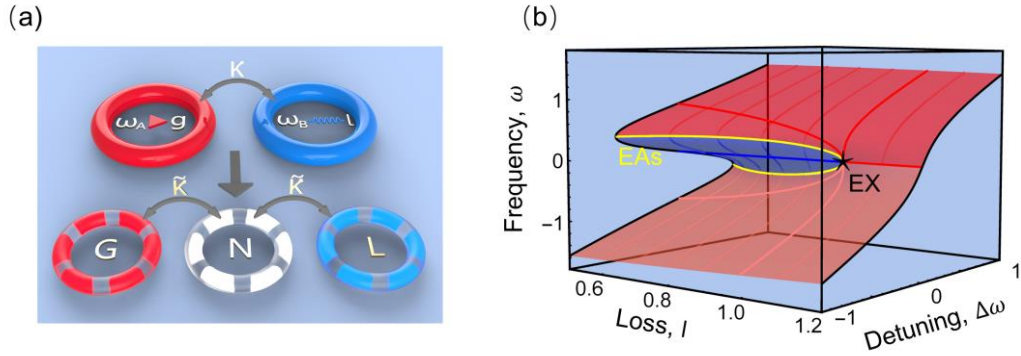

FIG. 1 (a) The upper panel shows two coupled resonators. The left resonator (red) has resonance frequency $\omega_A$ and nonlinear saturable gain, and the other resonator (blue) has resonance frequency $\omega_B$ and linear loss. Such a nonlinear non-Hermitian system can be mapped into a *PT* symmetric three-resonator system as shown in the lower panel with linear gain (G), neutral (N) and loss (L) resonators. (b) Steady-state solution of Eq (1) with $\kappa = 1$, $\omega_B = 0$ and all the other parameters normalized by $\kappa$. The red and blue regions represent stable and unstable states, respectively. The yellow lines represent EAs which are also boundaries of the stable and unstable states. The two yellow lines merge at an EX (black star), where two stable states and one unstable state coalesce.

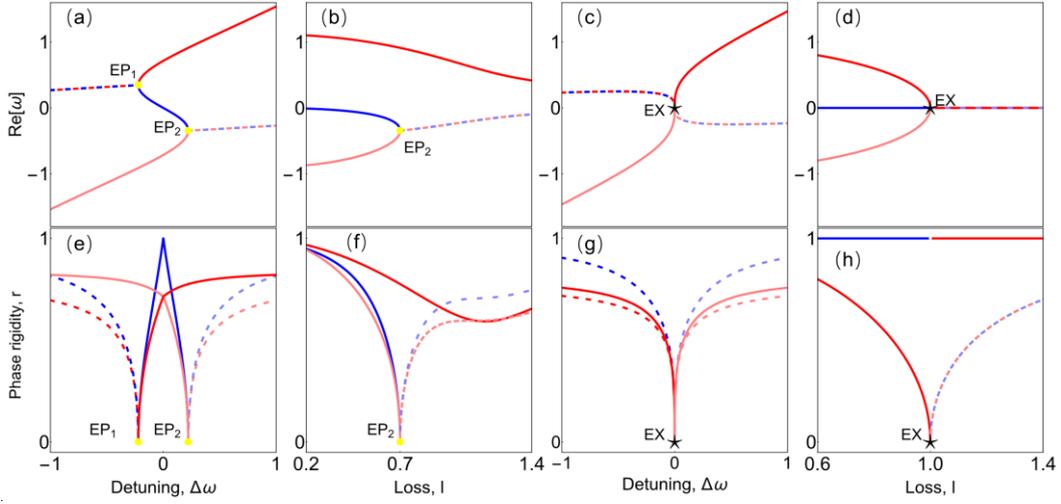

FIG. 2. (a)-(d) Real part of the solutions of Eq. (2) versus the detuning and loss, where the red and blue solid lines represent stable and unstable steady states, respectively, and the red and blue dashed lines represent complex solutions of Eq. (2) with positive and negative imaginary parts, respectively. $l = 0.7$ in (a), $\Delta_\omega = 0.22$ in (b), $l = 1$ in (c), $\Delta_\omega = 0$ in (d), and all the other parameters are the same as in Fig. 1(b). (e)-(h) Phase rigidities of the corresponding states in (a)-(d), respectively.

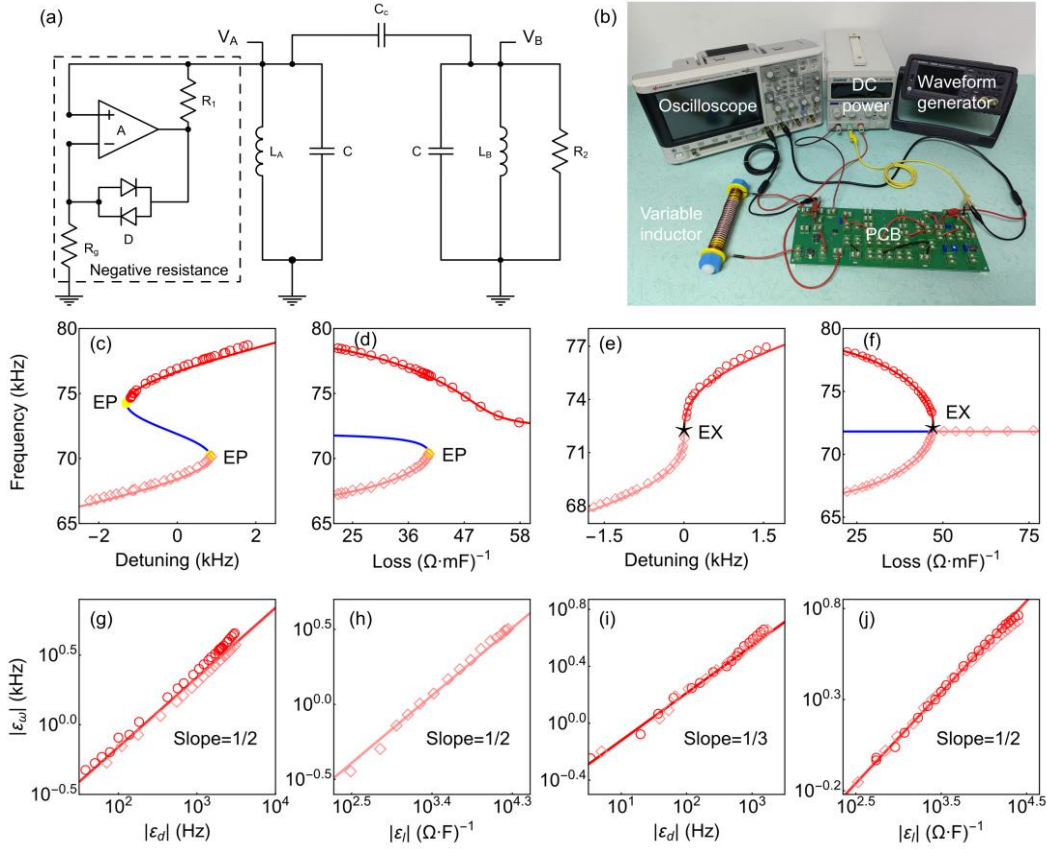

FIG. 3. (a) Circuit used for experimental verification, showing the inductors (L), capacitors (C), resistors (R), diodes (D) and amplifier (A). (b) Photo of the experimental setup. (c)-(f) Measured resonance frequencies (open circles and diamonds) of the system together with the steady state eigenfrequencies from the simulations (solid lines). Loss = 35.7 $(\Omega\,\text{mF})^{-1}$ in (c), detuning = 0.36(kHz) in (d), loss = 47.5$(\Omega\,\text{mF})^{-1}$ in (e), and detuning = 0(kHz) in (f). (g)-(j) show the critical behavior near the corresponding EPs and EX in (c)-(f), respectively. The solid lines are the theoretical predictions. The estimated experimental errors are smaller than the marker sizes, and are hence not explicitly shown.